# Reconfigurable optical-force-drive chirp and delay-line in micro/nano-fiber Bragg grating

Wei Luo,[1] Fei Xu,[1,2] and Yan-qing Lu[1]

[1] National Laboratory of Solid State Microstructures and College of Engineering and Applied Sciences, Nanjing University, Nanjing 210093, P. R. China
[2] E-mail: feixu@nju.edu.cn

**Abstract**
The emergence of optical micro/nano-fiber (MNF) with a subwavelength diameter, which has ultra-light mass and an intense light field, brings an opportunity for develop fiber based optomechanical systems. In this study, we theoretically show an optomechanical effect in silica MNF Bragg gratings (MNFBGs). The light-induced mechanical effect results in continuously distributed strain along the grating. It is shown that the power-related strain introduces an optically reconfigurable chirp in the grating period. We develop new optomechanical coupled-mode equations and theoretically analyze the influence of the optical-force-induced nonlinearity and chirp on the grating performance. Compared with weak Kerr effect, the optomechanics effect dominated in the properties evolution of MNFBGs and significant group velocity reduction and switching effect have been theoretically demonstrated at medium power level. This kind of optomechanical MNFBG with optically reconfigurable chirp may offer a path toward all-optical tunable bandwidth of Bragg resonance and may lead to useful applications such as all-optical switching and optically controlled dispersion and slow/fast light.

   **Introduction.** - Optomechanics is a rapidly developing field undergoing significant and continuous progress. Various theoretical models for and experiments into optomechanical effects have been proposed and demonstrated in various configurations [1-7]. In particular, transverse gradient optical forces and their mechanical Kerr effect have attracted increasing interest and have been extensively investigated in integrated photonic circuits based on nanoscale waveguides. In the mechanical Kerr effect, optical force can cause transverse deformations of waveguides and can change the effective index or coupling strength in some devices based on coupled free-standing nanoscale waveguides, such as resonators and interferometers. However, because of fiber size and device configuration, it is difficult to observe and utilize analogous optomechanical responses inside a conventional fiber system. Recent research into optomechanical effects in optical fibers mainly focused on one type of microstructured silica fiber: a dual-nanoweb structure suspended inside a capillary fiber. In dual-nanoweb structures, the strong optomechanical nonlinearity originating from a tight field confinement and a propagation length of several meters has been theoretically and experimentally investigated [8-10]. Nevertheless, the fabrication and implementation of related devices and their corresponding applications still present a significant challenge.
   Conversely, following a rapid development in fabrication technology, subwavelength-diameter optical micro/nano-fibers (MNFs) have attracted much attention because of their extraordinary advantages of large evanescence fields, light weight, and low-loss interconnection to standard fiber devices [11]. MNFs have emerged as an ideal basic element for miniature and functional fiber circuits or devices, including resonators and gratings. Indeed, in consideration of their extreme light weight and their free-standing status, MNFs also evince the influence of a weak optical force, which is characterized by the observable sideways displacement of the fiber tip due to an optical push force on the end face [3, 4]. Therefore, both enabling optomechanical effects and exploiting new applications in typical MNF devices such as gratings merit further attention.

Fiber gratings have become a critical component for many applications in fiber-optic systems. However, radiation pressure, originating from reflection in fiber Bragg gratings, has received less attention because this pressure has limited impact on the relatively large mass of conventional gratings. Even optical nonlinear behavior – such as optical bistability and self-switching [12-18] – is also difficult to observe under continuous-wave (CW) or quasi-CW excitation because the extremely high threshold power, which – without optimization – amounts to $\sim 10^4$ W. However, MNFs open new opportunities for research. With the emergence of low-loss MNFs, numerous techniques to inscribe gratings into MNFs have been successfully demonstrated over the past five years, for example, processing with an ultraviolet laser [19], a $CO_2$ laser [20], and a femtosecond laser [21]; and milling with a focused ion beam (FIB) [22]. With these gratings' very small cross-sectional area and their enhanced light intensity, the scale of radiation pressure is sufficient for creating a considerable longitudinal deformation in the structure, which thereby affects the optical characteristics of the grating.

Here we present a detailed analytical and theoretical model of light-induced chirp in a system of silica MNF Bragg gratings (MNFBGs) with air cladding. Chirped fiber Bragg gratings are of great importance for applications in optical communication and information processing. Chirping the period of a grating enables the dispersive properties of the scattered light to be tailored. Chirped fiber gratings are useful for dispersion compensation, for controlling and shaping short pulses in fiber lasers, and for creating stable CW and tunable mode-locked external-cavity semiconductor lasers [23]. Our theory and calculations show that optically reconfigurable chirp in the period of Bragg grating can be achieved and tuned by incident light. The strong optomechanical effect results in a remarkable change in the spectra of the grating. The optomechanical effect can be further enhanced by selecting a fiber with a low Young's modulus. This methodology represents an exciting new approach for achieving flexible all-optical tuning and for the reconfiguration of function-integrated fiber circuits. Reflection and transmission delay is also calculated, and this calculation shows that large tunable group velocity reduction can be obtained at the band edge on short-wavelength side of the Bragg resonance. Moreover, the switching capacity of MNFBGs may provide a new platform for additional optomechanical applications. Because of the similarity between fiber and planar devices, this theoretical investigation would prove helpful for waveguide gratings in planar integrated circuits.

**Theoretical model**. - Figure 1(a) shows the generic structure under study. The MNFBG is straight, with input end fixed and the other end free. The refractive index varies periodically along the z-axis. The index variation in the MNFBG is affected by both radiation pressure and Kerr nonlinearity, and this variation can be written as

$$n(z) = n_{eff} + n_a \cos\left(\frac{2\pi}{\Lambda[1+\varepsilon_e(z)]}z\right) + n_2|E(z)|^2 - n_{eff}p_{eff}\varepsilon_e(z) \quad (1)$$

where $n_{eff}$ is the fiber mode's effective refractive index, $n_a$ is the amplitude of the periodic index change, $\Lambda$ is the Bragg period, $n_2$ is the nonlinear Kerr coefficient, $E(z)$ is the electric field, and $p_{eff}$ is the effective photo-elastic coefficient. $\varepsilon_e(z)$ is continuously distributed strain caused by radiation pressure, which is given by

$$\varepsilon_e(z) = \frac{F_R(z)}{EA}, \quad (2)$$

where $E$ is the Young's modulus, while $A$ is the grating's cross-sectional area. The radiation pressure force $F_R(z)$ acting on the grating can be calculated by means of the Maxwell stress tensor:

$$F_R(z) = \int_{z=L} T_{zz} dS - \int_z T_{zz} dS \quad (3)$$

$$T_{zz} = \frac{1}{2}\varepsilon(E_z^2 - E_x^2 - E_y^2) + \frac{1}{2}\mu(H_z^2 - H_x^2 - H_y^2) \quad (4)$$

where $dS$ is the area element of the cross section. The optical force has the same direction as the light propagating along the z-axis. Note that $F_R(z)$ is nonuniform under the given mechanical boundary conditions, and thus $\varepsilon_e(z)$ is also nonuniform for MNFBGs. The nonuniform strain along the grating forms what we call light-induced chirp in the grating period.

Substituting Eq. (1) into the Maxwell equations, we obtain the following set of steady-state optomechanical coupled-mode equations without loss and dispersion:

$$i\frac{\partial A_f}{\partial z} + \delta A_f + \kappa A_b + \gamma(|A_f|^2 + 2|A_b|^2)A_f$$
$$+ \eta(|A_f|^2 + |A_b|^2 - |A_f(L)|^2)A_f - \frac{1}{2}\frac{\partial \phi}{\partial z}A_f = 0 \quad (5)$$

$$-i\frac{\partial A_b}{\partial z} + \delta A_b + \kappa A_f + \gamma(|A_b|^2 + 2|A_f|^2)A_b$$
$$+ \eta(|A_f|^2 + |A_b|^2 - |A_f(L)|^2)A_b - \frac{1}{2}\frac{\partial \phi}{\partial z}A_b = 0 \quad (6)$$

$$\gamma = \frac{n_2 k_0}{A_{eff}} \quad (7)$$

$$\eta = \frac{1}{2}n_{eff}p_{eff}k_0 \frac{\int\left[\frac{1}{2}\varepsilon(e_z^2 - e_x^2 - e_y^2) + \frac{1}{2}\mu(h_z^2 - h_x^2 - h_y^2)\right]dS}{EA} \quad (8)$$

$$\phi = \frac{\pi}{\Lambda} z \frac{\int \left[ \frac{1}{2}\varepsilon(e_z^2 - e_x^2 - e_y^2) + \frac{1}{2}\mu(h_z^2 - h_x^2 - h_y^2) \right] dS}{EA} (|A_f|^2 + |A_b|^2 - |A_f(L)|^2) \quad (9)$$

where $A_f$ and $A_b$ are the envelope functions of the forward- and backward-traveling waves, $\delta = \beta - \beta_B$ is the detuning parameter, $\beta$ is the propagation constant, $\beta_B = \pi/\Lambda$ is the Bragg wave number, $\kappa = \pi n_a / \lambda$ is the coupling coefficient and $(1/2) d\phi / dz$ describes light-induced chirp of the grating period (considering the strain in grating is usually very small, the expression for $\phi$ has certain approximation). $\gamma$ represents the nonlinear parameter, which is inversely proportional to the effective mode area $A_{eff}$. This relation indicates that MNFBGs with a small effective mode area may have a large nonlinear parameter. The difference between these equations and previous theory lies in the terms related to the time-average optomechanical parameter $\eta$ and optomechanical chirp. In their definitions, the mode field distributions, $e_i$ and $h_i$ ($i = x, y, z$), are already normalized by the power flow. Equations (5) and (6) are valid for CW optical input at wavelengths near the Bragg wavelength, but the steady-state solutions in the quasi-CW regime are close to the solutions in the CW regime [16]. Therefore, for simplicity, we only consider the case under CW excitation in the fundamental mode.

Figure 1(b) summarizes the nonlinear and optomechanical parameters of silica MNFBGs at different radii. The operating wavelength of the monochromatic light is 1550 nm, the nonlinear Kerr coefficient is $2.6 \times 10^{-20}$ m$^2$/W, the effective photo-elastic coefficient is 0.21, Young's modulus is 70 GPa, and the Bragg wavelength is set to 1550 nm. The refractive index of silica glass is determined by the Sellmeier polynomial. As we can see in the figure, nonlinear parameter is positive and optomechanical parameter is negative in MNFBGs with radii ranging from 300 nm to 1.5 μm, because Kerr effect and photo-elastic effect have opposite impact on the index change in the grating. The nonlinear parameter increases but the optomechanical parameter decreases the effective index of the fiber material. They are both much larger than that in standard fiber gratings with a diameter of 125 μm. Absolute values of both parameters continue to increase with decreasing radius until these parameters reach their peaks. This result arises because light can no longer be tightly confined in MNFs with smaller radii.

**Results**. - Solutions for the steady-state optomechanical coupled-mode equations can be obtained by a semi-implicit Runge-Kutta method and iteration. Figure 2 shows the calculated spectra using these methods. The wavelength of input light is around 1550 nm, with a grating length of 1 cm and a coupling coefficient of 2 cm$^{-1}$. We calculate and plot the transmission spectra of an MNFBG with a 300-nm radius at four different output powers in Fig. 2. The power is normalized by the critical power, $P_{cr} = 4/(3\gamma L)$ [12].

At low power level, the transmission spectrum is symmetrical (shown in Fig. 2, black line). With the increasing of the power, the shape of the spectra eventually becomes asymmetric and the width of dips increases, and the minimum transmittance increases at the same time. These are consequences of Kerr effect and optomechanical effects (including photo-elastic effect and chirp induced by radiation pressure). Radiation pressure plays a major role among all these effects. Chirp in the grating period introduces shift of Bragg wavelength, which is the main reason of bandwidth broadened and transmittance increasement. We can get broader reflection bandwidth by higher incident laser power below laser-induced damage threshold.

**Optically reconfigurable chirp**. - We investigate the internal strain distribution produced by optical force in the 300-nm-radius MNFBG at normalized input power ranging from 0 to 0.4. The wavelength of input light is 1550 nm and 1549.90 nm, respectively. Strain distribution in the grating is plotted in Fig. 3 (a) and (b). At wavelength of 1550 nm, strain induced by radiation pressure shows a declining trend along the length since the light wave in the photonic bandgap decays exponentially. The distribution remains unchanged with the power level, but strain at high power gets larger than that at low power. Therefore, the magnitude of the strain can be controlled by the input light. It gives us the possibility to achieve an optically reconfigurable chirp in the grating period, and is promising to open up a new platform for more optomechanical applications involving all-optical tunable filter. At wavelength of 1549.90 nm, the result has changed. The maximum strain does not locate in the beginning of the grating at high power. Hence, strain distribution can also be tuned by the wavelength of the input light. Finally, unlike the optical Kerr effect, the optomechanical effect can be transmitted and can produce an evident influence on the external fiber outside the MNFBG. Light-actived stress in the MNFBG can be used to mechanically modulate some force-sensitive elements (slot-MNF, grating or coupler) in the external MNF. It is possible to have more optomechanical applications involving interactions between interconnected MNF function elements, which cannot be realized by pure optical nonlinearity.

**Group velocity reduction**. - Delay characteristics of the system are shown in Fig. 4. As is displayed in the figures, a large delay can be obtained at the band edge on short-wavelength side of the Bragg resonance. Thus we can get a notable group velocity reduction in optomechanical MNFBGs. The maximum reflection delay and transmission delay can reach about 400 and 225 ps, respectively. These delays are equivalent to group velocity of about $2.5 \times 10^7$ and $4.4 \times 10^7$ m/s, respectively. In contrast, delay at the band edge only reaches ~72 ps in the standard 125-μm-diameter fiber grating with the same structural parameters [23]. The emergence of the light-induced chirp also change the situation in nonlinear Bragg gratings. If we just consider the Kerr nonlinearity in the grating, the larger delay should appear at the band edge on long-wavelength side. The chirp leads to an opposite result. Moreover, the group velocity reduction is enhanced by the additional chirp in optomechanical MNFBGs. According to our calculation results, wavelength of the band edge and delay can both be tuned by input light. Larger delay at shorter wavelength is possible by increasing the incident power.

**Switching capacity**. - All-optical switching in MNFBGs can be achieved at certain incident wavelength. We plot the relation between normalized input power and transmittance of the 300-nm-radius MNFBG in Fig. 5. As is seen in the curves, transmittance of the grating decreases with the incident power at wavelengths of 1549.84, 1549.87 and 1549.90 nm, but remains almost unchanged when the input wavelength is set to 1550.00 nm. Different from optical self-switching in conventional nonlinear grating, switching in optomechanical MNFBGs does not contain bistability, but it has sufficient extinction ratio at high input power. Other than self-switching, switching by the cross-phase modulation can also be realized, which is induced by an intense pump pulse on a low intensity probe.

**Discussion.** - So far, we have discussed the static response of the system. Dynamic response is another interesting topic. The optomechanical effects are always much slower than the electronic Kerr effect in MNFBGs. The speed of the optomechanical response is limited by the mechanical resonant frequencies of the structure. Behaviors at ultra-high frequency may be quite different for optomechanical and Kerr effects, that is, the high-frequency modulation may seem quasi-continuous for optomechanical response, but still discontinuous for the response of Kerr nonlinearity. As a consequence, the optomechanical effect may exhibit a steady-state behavior while the Kerr effect shows a dynamic response. This possible characteristic may pave a way for new applications of the optomechanical devices. Moreover, the optomechanical response can be enhanced by excitation at the mechanical resonant frequency.

In addition, the interplay of Kerr nonlinearity and dispersion in Bragg gratings, as is known, leads to the formation and propagation of gap solitons [24, 25]. If we consider the dispersion in MNFBGs, the optomechanical effect will surely have influence on soliton formation and propagation. There is no space to cover them in depth here, but it is worth our attention to ponder these possibilities.

In conclusion, we have theoretically demonstrated that silica MNFBGs exhibit strong optomechanical behavior, and have furthermore obtained numerical solutions for these gratings. The light-induced strains introduce additional period and refractive index changes to the Bragg condition for coupling between forward and backward waves within the gratings, which result in a tunable bandwidth of the Bragg reflection. The chirp generated by radiation pressure can be optically activated and tuned by modulating the input light. This optically reconfigurable chirp indicates an approach for all-optically manipulating MNF devices. The results of group velocity reduction and switching capability of MNFBGs may permit potential applications, such as tunable optical delay line and all-optical switching. Finally, this theoretical analysis will likely prove helpful for waveguide gratings in planar integrated circuits, including silicon-on-insulator on-chip systems, for which previous studies have mainly focused on the transverse gradient force.

**Acknowledgements**
This work is supported by National 973 program under contract No. 2012CB921803 and 2011CBA00205, National Science Fund for Excellent Young Scientists Fund (61322503) and National Science Fund for Distinguished Young Scholars (61225026). The authors thank Prof. Xiaoshun Jiang and Xuejing Zhang for help on numerical simulation.

Figure Captions:

FIG. 1. (a) Schematic of the MNFBG structure. The notation used in the letter is also illustrated. (b) Values for nonlinear and optomechanical parameters of silica MNFBGs at different radii. Black squares represent the nonlinear parameter and red dots represent the optomechanical parameter. Black solid line and red solid line represent the fitted curves.

FIG. 2. Transmission spectra of 300-nm-radius MNFBG with normalized output power of $5\times10^{-5}$, $5\times10^{-3}$, $4.5\times10^{-2}$ and 0.125.

FIG. 3. (a) Time-averaged strain distribution in the MNFBG with a 300-nm radius at incident wavelength of 1550.00 nm. (b) Time-averaged strain distribution in the MNFBG with a 300-nm radius at incident wavelength of 1549.90 nm.

FIG. 4. (a) Reflection delay of 300-nm-radius MNFBG at normalized output power of $4.5\times10^{-2}$. (b) Transmission delay of 300-nm-radius MNFBG at normalized output power of $4.5\times10^{-2}$.

FIG. 5. Relation between normalized incident power and transmittance of 300-nm-radius MNFBG at wavelength of 1549.84, 1549.87, 1549.90 and 1550.00 nm.

Figure 1:

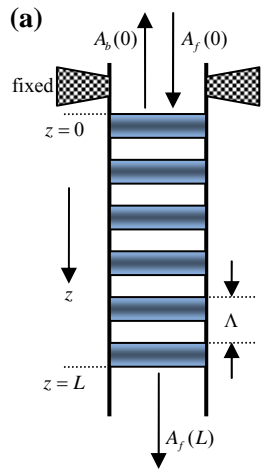 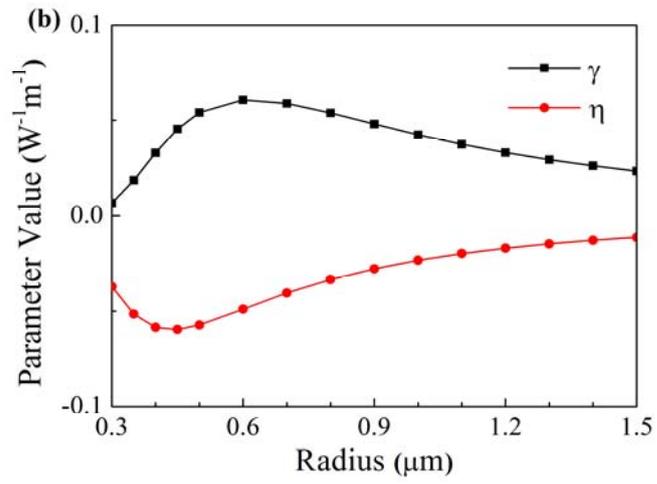

Figure 2:

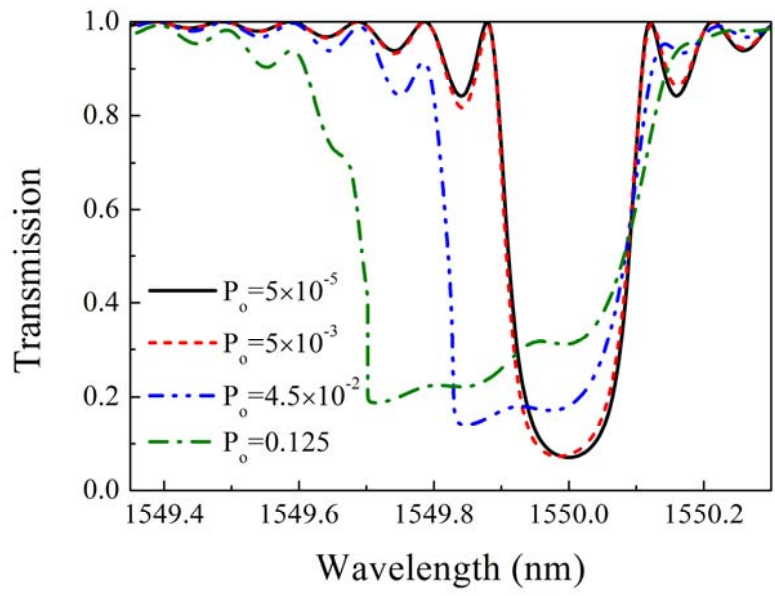

Figure 3:

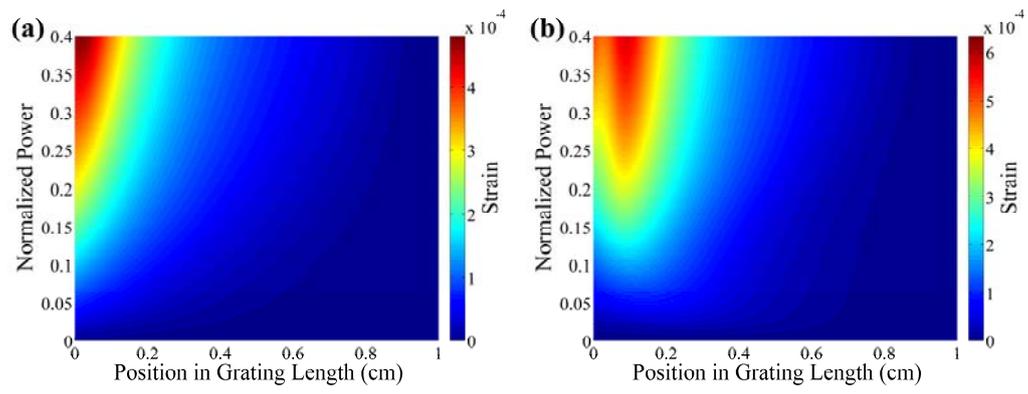

Figure 4:

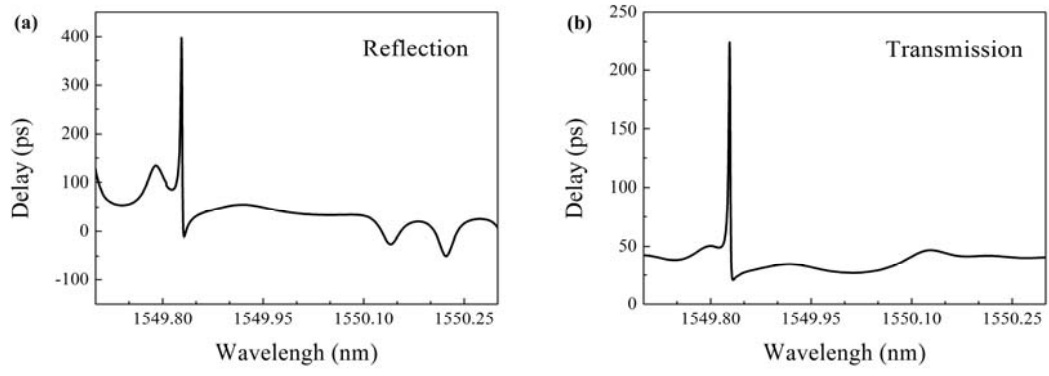

Figure 5:

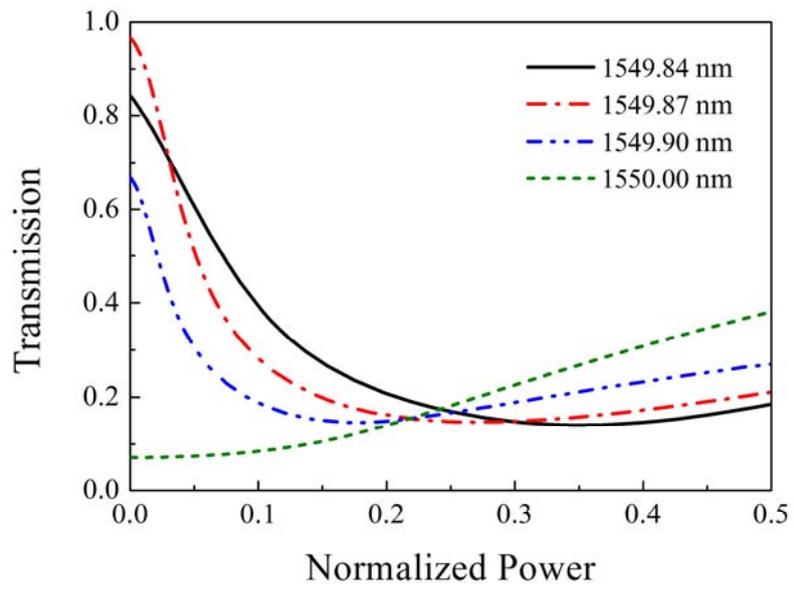